# Characterization of Coulomb Interactions in Electron Transport through a Single Hetero-Helicene Molecular Junction Using Scanning Tunneling Microscopy

Yueqing Shi[1*], Liya Bi[1,2*], Zihao Wang[1,2*], Kangkai Liang[1,2], Ji-Kun Li[3], Xiao-Ye Wang[3], Wan-Lu Li[2,4], and Shaowei Li[1,2†]

[1] Department of Chemistry and Biochemistry, University of California, San Diego, La Jolla, CA 92093-0309, USA

[2] Program in Materials Science and Engineering, University of California, San Diego, La Jolla, CA 92093-0418, USA

[3] State Key Laboratory of Elemento-Organic Chemistry, College of Chemistry, Nankai University, Tianjin, 300071, China

[4] Aiiso Yufeng Li Family Department of Chemical and Nano Engineering, University of California, San Diego, La Jolla, CA 92093-0448, USA

**Abstract**

Characterization of the structural and electron transport properties of single chiral molecules provides critical insights into the interplay between their electronic structure and electrochemical environments, providing broader implications given the significance of molecular chirality in chiroptical applications and pharmaceutical sciences. Here, we examined the topographic and electronic features of a recently developed chiral molecule, B,N-embedded double hetero[7]helicene, at the edge of Cu(100)-supported NaCl thin film with scanning tunneling microscopy and spectroscopy. An electron transport energy gap of 3.2 eV is measured, which is significantly larger than the energy difference between the highest occupied and the lowest unoccupied molecular orbitals given by theoretical calculations or optical measurements. Through first-principles calculations, we demonstrated that this energy discrepancy results from the Coulomb interaction between the tunneling electron and the molecule's electrons. This occurs in electron transport processes when the molecule is well decoupled from the electrodes by the insulating decoupling layers, leading to a temporary ionization of the molecule during electron tunneling. Beyond revealing properties concerning a specific molecule, our findings underscore the key role of Coulomb interactions in modulating electron transport in molecular junctions, providing insights into the interpretation of scanning tunneling spectroscopy features of molecules decoupled by insulating layers.

---

* These authors contribute equally to this work.

† To whom all correspondence should be addressed: shaoweili@ucsd.edu.

**Introduction**

Chiral luminescent materials have broad applications in optoelectronic devices such as chiral sensors and organic light-emitting diodes[1-4]. Among these materials, helicenes, a class of intrinsically chiral polycyclic aromatic hydrocarbons, have attracted considerable attention for their exceptional chiroptical properties owing to their uniquely distorted helical geometry and extended π-conjugation[5, 6]. Recently synthesized B,N-embedded double hetero[7]helicenes (DH) have demonstrated impressive chiroptical properties[7], including strong chiroptical activities from 300 nm to 700 nm and efficient circularly polarized luminescence from 600 nm to 800 nm. These excellent properties make them promising candidates for chiral optoelectronic applications in the visible and near-infrared regions. Nevertheless, this prospect is gapped by the absence of knowledge on the structural and electronic properties of DH molecules at the molecular scale.

Scanning tunneling microscope (STM) is a powerful tool for investigating the structural and electronic properties of single molecules on surfaces. It directly images the morphology and adsorption state of molecules on different surfaces[8-10]. Additionally, it can correlate the local electronic properties to the nanoscale structural features via measuring the density of states (DOS) with scanning tunneling spectroscopy (STS), which has been demonstrated for metal[11-13], semiconductor[14, 15] and molecular[16, 17] systems. When directly adsorbed on metal surfaces, the electronic structures of the molecules are usually highly distorted by the electron sea underneath. However, it has been shown that the intrinsic molecular orbitals of single surface-adsorbed molecules can be preserved and surveyed with STS when an insulating film such as NaCl is added between the molecule and metal substrate[18-24]. This offers an exciting platform to gain insights into the electron transport properties of the single-molecule junction consisting of a DH molecule.

In this study, we investigated individual DH molecules on both the Cu(100) surface and at the NaCl edge combining low-temperature STM measurement and density-functional theory (DFT) calculations. We characterized the electron transport from and to the frontier molecular orbitals of the DH molecule adsorbed at the NaCl edge with STS and found the measured energy gap between the highest occupied molecular orbital (HOMO) and the lowest unoccupied molecular orbital (LUMO) to be ~1.3-1.4 eV larger than the optical gap revealed by ultraviolet-visible (UV-Vis) absorption or Photoluminescence (PL) measurements. Based on our DFT calculations, we assign this difference to the temporary ionization of the molecule by the tunneling electron, which increases the measured transport gap to account for the additional Coulomb repulsion/attraction energy required for an electron to tunnel into/out from the molecular orbitals. This additional energy is only observed when the molecule is strongly decoupled from the Cu surface by a NaCl thin film and is absent when the molecule is more weakly decoupled, such as when bound to a defect site on the NaCl island.

## Results and Discussion

The top panels in **Figure 1a** depict the molecular structures of both enantiomers, namely, left-handed (L-) and right-handed (R-) DH molecules. The racemates of DH molecules were sublimated (**Figure S1**) onto the substrate surface in the ultra-high vacuum (UHV) chamber as detailed in **Supporting Information**. Shown in **Figure 1b** is the STM image of a pair of mirror-symmetric molecules on the bare Cu(100) surface. We assigned the molecule that has a contour resembling the letter *S* to L-DH (left in **Figure 1b**) and the other one to R-DH (right in **Figure 1b**) according to their 3D structures in **Figure 1a**. In other words, when adsorbed on Cu(100), the DH molecules mainly adopt a geometry that is overall parallel to the surface (**Figure 1a,** bottom panels), as supported by the DFT-calculated adsorption geometry of an R-DH molecule on Cu(100) (**Figure S4**). The fact that DH molecules are stably adsorbed indicates a substantial orbital hybridization with the metal substrate. This is supported by the STS measurements showing that the DOS of an R-DH molecule adsorbed on Cu(100) is nearly identical to that of the substrate (**Figure 1c**).

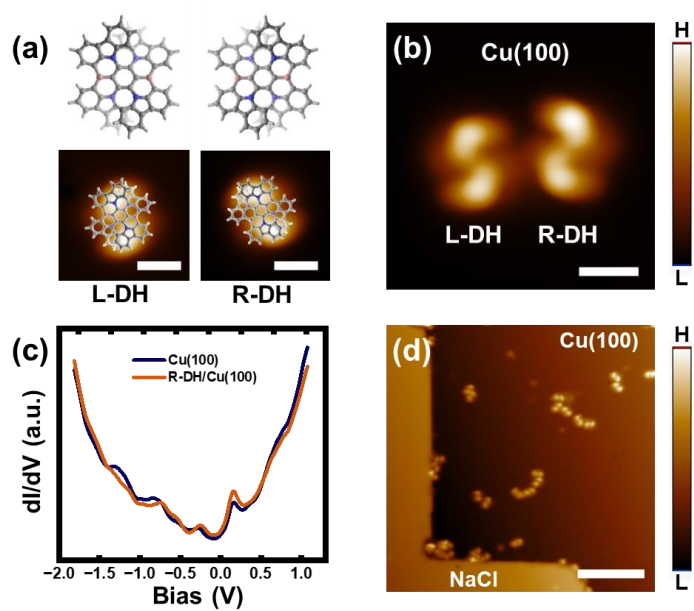

**Figure 1.** Adsorption of DH molecules on Cu (100). (**a**) Top panels: top view of the 3D structures of the DH molecules with both chirality, i.e., L- and R-. The C, N, and B atoms are shown in grey, blue, and red. The color gradient represents the relative vertical positions of the atoms. Darker color corresponds to higher *z*. Bottom panels: STM images of individual DH molecules with their molecular structures overlaid. Scale bar=1 nm. (**b**) Topographic image of a pair of L- and R-DH molecules adsorbed on Cu(100) (scale bar=1 nm). Imaging parameters were set to 500 mV, 30 pA. (**c**) STS spectra of an R-DH molecule on Cu(100) (orange) and the substrate (blue). The tunneling gaps were set to -2 V and 3 nA. (**d**) Large-area scan of the DH molecules on surfaces. Scale bar=10 nm. Imaging parameters were set to 500 mV, 20 pA.

Thin-layer NaCl has been widely utilized to disentangle the molecules from the metal substrates for characterizing the molecular electronic structure with STS[18, 38, 39]. When inspecting the surface, we found that DH molecules predominantly stayed on the bare Cu(100) areas and occasionally came across DH molecules that were adsorbed at the edges of NaCl thin films (**Figure**

**1d** and **Figure 2a**), but never observed a DH molecule on the NaCl terraces. This finding indicates negligible binding of DH molecules to the NaCl/Cu(100) surface, which leads to unstable molecular adsorption even at such a low temperature. Our DFT calculations well reproduced this and revealed an adsorption energy of -1.04 eV when an R-DH molecule binds to Cu(100), as compared to only -0.28 eV on NaCl/Cu(100) (**Figure S4**). To investigate the electronic properties of DH molecules, we focused on the DH molecules that resided at the NaCl edges. **Figure 2b** shows the zoom-in image of such an R-DH molecule, whose chirality was confirmed by pushing it to the adjacent Cu(100) surface with the STM tip after all measurements (**Figure S5**). According to the DFT calculations, the R-DH molecule covers the NaCl edge with an ~ 2.5 Å gap between the lowest H atom and the surface Cu atom (**Figure 2c**, left).

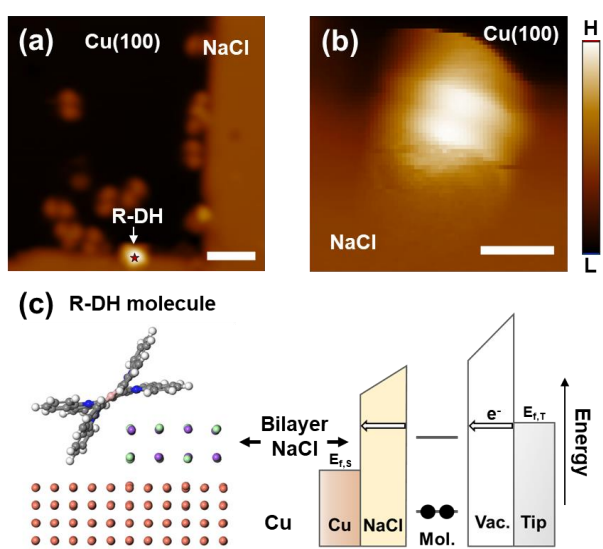

**Figure 2.** Adsorption of DH molecules at the edge of NaCl/Cu(100). (**a**) STM image of DH molecules on Cu(100) and an R-DH molecule at the NaCl edge. Scale bar=3 nm. (**b**) Zoom-in image of the R-DH molecule in (a). Scale bar=1 nm. Imaging parameters were set to (a) 800 mV, 7 pA and (b) 1.65 V, 7 pA. (**c**) Left: side view of the DFT-calculated adsorption geometry of an R-DH molecule at the NaCl edge. Right: schematic representation of electron transport through the adsorbed DH molecule in the double-barrier junction at positive bias.

When sandwiched in between the STM tip and NaCl/Cu(100) surface, the DH molecule is effectively in a double-barrier tunnel junction as sketched in the right panel of **Figure 2c**. **Figure 3a** showcases a representative STS spectrum taken over the R-DH molecule in **Figure 2b**. A HOMO-LUMO gap of ~ 3.2 eV was clearly resolved and is consistent with the effective decoupling of the molecule from Cu(100) surface as captured by the theory (**Figure 2c**, left). No obvious difference in electronic structure is resolved between molecules with different chirality. Small variation in spectrum features in dI/dV is observed among molecules adsorbed at different sites, likely due to the inhomogeneity in local chemical environment. Notably, the HOMO-LUMO gap of the single surface-adsorbed DH molecule measured with STS is significantly larger than the optical gap, ~1.87-1.97 eV, as deduced from the UV-Vis (**Figure 3b**) or PL (**Figure 3c**) spectra of racemic

DH molecules dissolved in toluene. It is also worth noting that the HOMO in the dI/dV spectrum exhibits a double peak with an energy separation of approximately 150 meV. This feature, while captured in UV-Vis (**Figure 3b**) and PL (**Figure 3c**) spectra, was not captured by our DFT calculations (**Figure 4a**). This double peak may result from the further broken of molecular inversion symmetry upon surface adsorption, in conjunction with the lifting of spin degeneracy in HOMO due to spin-orbit coupling, a phenomenon that is challenging to simulate accurately with the DFT package we employed.

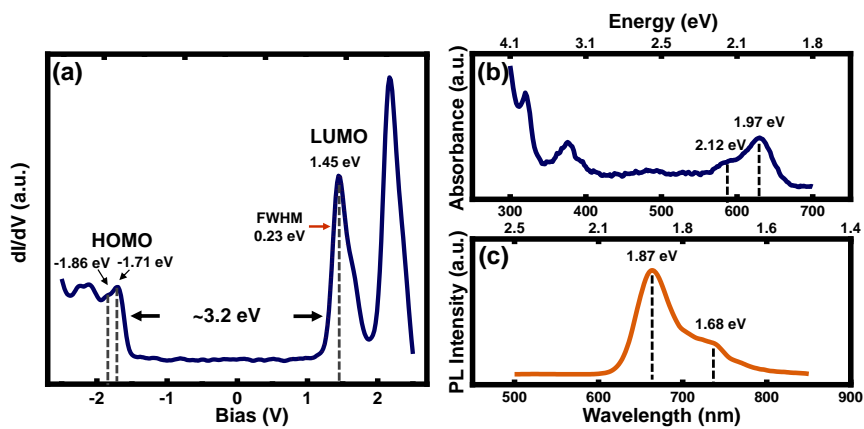

**Figure 3.** Characterizing the electronic structures of DH molecules. (**a**) STS spectrum of an R-DH molecule at the NaCl edge. The tunneling gap was set to 2.5 V and 100 pA. The red dotted line is a Gaussian fit for LUMO. (**b**) UV-vis absorption and (**c**) PL spectra of the racemic DH molecules in toluene.

To explain the significant difference between the HOMO-LUMO gap observed in STS and optical studies, it is important to consider the temporary ionization of the molecule during electron transport which is absent in optical excitation processes. Usually, the measured gaps from both electron transport and optical excitation are deviated from the intrinsic energy difference between HOMO and LUMO. The gap observed in STS could be wider than the actual gap because it reflects the transiently charged state of the molecule, requiring consideration of the internal electrostatic potential[20, 23]. In contrast, the optical gap could be narrower than the fundamental gap owing to the attractive Coulomb interaction within the electron-hole pair or the exciton binding energy upon optically exciting the molecules. Therefore, both the widening of the HOMO-LUMO gap and the narrowing of the optical gap could contribute to the observed ~1.3-1.4 eV difference between them. However, we propose that the broadened HOMO-LUMO gap is the primary contributor to this significant energy difference since the exciton binding energy in most helicenes has been reported to be less than 0.5 eV [40-42]. Specifically, when measuring the molecular orbitals on NaCl/Cu(100) with STS, the molecule is temporarily charged due to the weak hybridization with the metal substrate with the insulating layer in between, leading to a prolonged lifetime of the charged states up to tens of nanoseconds[39]. The localized electron/hole on the LUMO/HOMO for a closed shell molecule like DH will thus transiently lift/lower the corresponding orbital because of the Coulomb energy involved in adding/removing an electron to/from the molecule, leading to an enlarged HOMO-LUMO gap.

This proposed mechanism is corroborated by further DFT calculations. The calculated DOS of a neutral R-DH molecule at the NaCl edge reveals a HOMO-LUMO gap of 1.45 eV (**Figure 4a**), which is only slightly larger than that of a gas-phase molecule (**Figure S6**), indicating negligible orbital perturbation from both NaCl and Cu. To accurately model the electron transport processes within the junction during STS measurement, the ionization of the molecule is incorporated into calculations, which include the interaction terms for electron tunneling into or out of the molecular orbitals. The energy level for LUMO tunneling is computed using self-consistent charge density, accounting for repulsive Coulomb interactions between the tunneling electron (into LUMO) and the occupied orbitals. A similar analysis is applied to the HOMO, where the interaction term becomes attractive after electron tunneling. The HOMO-LUMO gap is determined through partial-occupancy calculations with fixed self-consistent wave functions. After factoring in Coulomb repulsion and attraction during tunneling, the gap enlarges by 1.68 eV resulting from shifts in the LUMO and HOMO (**Figure 4b**), aligning with the ~1.3-1.4 eV difference between the STS and optical gaps of the DH molecule.

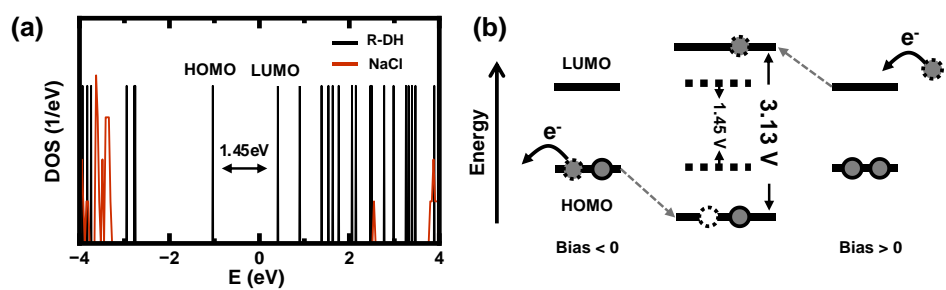

**Figure 4.** Computational insights into the HOMO-LUMO gap of single surface-adsorbed R-DH molecule during electron transport. (**a**) DFT-calculated DOS of the R-DH molecule at the NaCl edge (black) and the bilayer NaCl (red). (**b**) Schematics showing the broadened HOMO-LUMO gap after considering the on-site Coulomb interactions due to the temporary charging of an R-DH molecule in the STS measurements.

Additionally, the enlargement of the HOMO-LUMO gap in STS measurement is absent when the molecule is bound to a defect site at the edge of the NaCl island (**Figure 5**, left inset), where the molecule-substrate hybridization is stronger than at the intact NaCl edge (**Figure 2b,c**) but still much weaker than for molecules directly adsorbed on the Cu surface. In this case, STS resolves the LUMO and HOMO at 1.15 eV and -1.45 eV, respectively, corresponding to a ~2.6 eV gap, which is much closer to that derived from the optical measurements as compared to the gap for molecules adsorbed at the intact NaCl edge. The weak conductance features observed between -1.2 and 0 V bias are assigned to the Cu surface states by comparing with the dI/dV of bare Cu surface (**Figure 1c**). Besides, both the LUMO and HOMO peaks in the STS measured at the defect site are broadened, with long tails in the dI/dV signal extending across the Fermi level. The full width at half maximum (FWHM) for the LUMO peak of the molecule at the intact NaCl edge is ~ 0.23 eV (**Figure 3a**), about half that of the molecule bound to the defect (**Figure 5**). This broadening results from the hybridization of discrete molecular orbitals with the continuum metal states of the substrate, creating a tunneling channel through which the incoming electron quickly leaves the molecule, exchanging energy directly with the Cu lattice. As a result, the Coulomb interaction does not obviously influence the tunneling process.

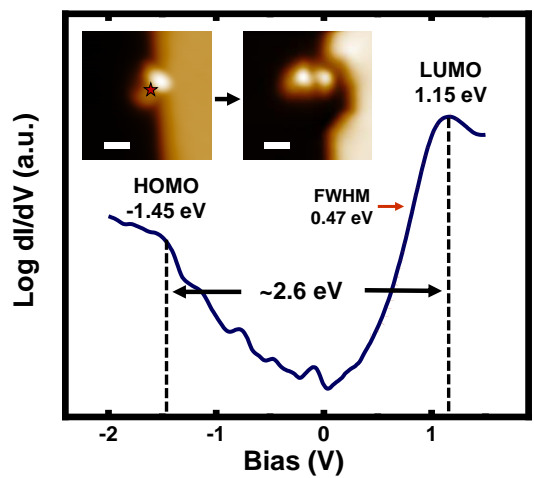

**Figure 5.** STS spectrum of an L-DH molecule at the NaCl edge defect (insets) plotted in the log scale. The red dotted line is a Gaussian fit for LUMO. The tunneling gap was set to -2 V and 300 pA. The right inset shows the same molecule in the left one after being pushed to Cu(100) by the STM tip. Scale bar = 1 nm. Imaging parameters were set to -1.14 V, 30 pA (left inset) and -1 V, 30 pA (right inset). The small dI/dV features between -1.2 and 0.5 V are also observed in the STS of bare Cu surface, and are assigned as the surface states of Cu.

## Conclusions

In this study, we characterized the structural and electronic properties of a single DH molecule in a double-barrier tunneling junction formed between the STM tip and a NaCl thin film supported by Cu(100) substrate. The chirality of the molecule is easily identifiable in topographic images when it is adsorbed directly on Cu(100). We found that DH molecules tend to avoid adsorption on the terrace of bilayer NaCl but can weakly bond to the edge of the NaCl island. The weak molecule-bath interaction extends the lifetime of the charged states, leading to the temporary ionization of the molecule during the STS measurement. Consequently, the measured HOMO-LUMO gap in STS is significantly broader for the molecules at the NaCl edge due to the Coulomb interaction that the tunneling electrons experience when passing through the molecule. This effect is absent in the molecule adsorbed on the detect sites of NaCl, where the molecule-substrate hybridization is enhanced. Our study provides molecular-scale insights into the electron transport properties of single DH molecules in an STM junction and highlights how they are influenced by molecule-substrate hybridization.

**Acknowledgments**

This work was supported by the United States National Science Foundation (NSF) under Grant No. CHE-2303936 (to Shaowei Li). The authors acknowledge the use of facilities and instrumentation supported by NSF through the UC San Diego Materials Research Science and Engineering Center (UCSD MRSEC) with Grant No. DMR-2011924. This work used the San Diego Supercomputer Center (SDSC) Expanse at UC San Diego for theoretical calculations through allocation CHE-240050 from the Advanced Cyberinfrastructure Coordination Ecosystem: Services & Support (ACCESS) program, which is supported by NSF Grants No. OAC-2138259, No. OAC-2138286, No. OAC-2138307, No. OAC-2137603, and No. OAC-2138296.